\documentclass[journal,doublespacing]{IEEEtran}
\usepackage[latin9]{inputenc}
\usepackage{color}
\usepackage{float}
\usepackage{textcomp}
\usepackage{mathrsfs}
\usepackage{amsmath}
\usepackage{amssymb}
\usepackage{graphicx}
\onecolumn

\makeatletter

\usepackage{caption}
\usepackage{float}
\usepackage{setspace}
\usepackage{graphicx}
\usepackage{graphicx,dblfloatfix}
\usepackage{tikz}
\usepackage{color}
\usetikzlibrary{shapes,arrows,shadows}
\ifCLASSINFOpdf
\else
\fi
\usepackage{capt-of}

\usepackage{caption}
\makeatother

\begin{document}

\title{PDL Impact on Linearly Coded Digital Phase Conjugation Techniques
in CO-OFDM Systems}

\author{O. S. Sunish Kumar, A. Amari, O. A. Dobre, R. Venkatesan\thanks{The authors are with the Faculty of Engineering and Applied Science,
Memorial University, St. John\textquoteright s, NL A1B 3X5, Canada
(e-mail: skos71@mun.ca). }
}
\maketitle
\begin{abstract}
We investigate the impact of polarization-dependent loss (PDL) on
the linearly coded digital phase conjugation (DPC) techniques in coherent
optical orthogonal frequency division multiplexing (CO-OFDM) superchannel
systems. We consider two DPC approaches: one uses orthogonal polarizations
to transmit the linearly coded signal and its phase conjugate, while
the other uses two orthogonal time slots of the same polarization.
We compare the performances of these DPC approaches by considering
both aligned- and statistical-PDL models. The investigation with aligned-PDL
model indicates that the latter approach is more tolerant to PDL-induced
distortions when compared to the former. Furthermore, the study using
statistical-PDL model shows that the outage probability of the latter
approach tends to zero at a root mean square PDL value of 3.6 dB.
On the other hand, the former shows an outage probability of 0.63
for the same PDL value.
\end{abstract}

\begin{IEEEkeywords}
Digital phase conjugation (DPC), fiber Kerr nonlinearity, polarization-dependent
loss (PDL). \vspace{-0.28cm}
\end{IEEEkeywords}

\section{Introduction}

\IEEEPARstart{T}{he} fiber Kerr nonlinearity and its interplay with
the random polarization effects limit the achievable capacity of the
polarization division multiplexed (PDM) coherent optical orthogonal
frequency division multiplexing (CO-OFDM) systems {[}1{]}. Over the
last decade, there have been extensive efforts to surpass the Kerr
nonlinearity limit through the development of several digital signal
processing techniques {[}2{]}. The most investigated digital nonlinear
compensation (NLC) techniques are digital back propagation, Volterra
series-based nonlinear equalization, and perturbation-based methods.
However, the large computational complexity limits the practical implementation
of such techniques in coherent optical systems {[}2{]}. In {[}3{]},
a digital phase conjugation (DPC) based technique, referred to as
the phase conjugated twin wave (PCTW), is proposed for the mitigation
of the first-order nonlinear distortions in PDM optical transmission
systems. The PCTW technique can be implemented with minimal additional
digital signal processing, providing a simple and effective solution
for optical fiber nonlinearity mitigation {[}4{]}. However, this gain
comes with halving the spectral efficiency of the PDM coherent optical
system {[}5{]}. Recently, in {[}6{]}, we proposed a linear polarization-coded
phase conjugated twin signals (LPC-PCTS) technique to solve the spectral
efficiency issue of the PCTW technique. In this scheme, the data symbols
on the adjacent subcarriers of the OFDM symbol are linearly coded
and transmitted as phase conjugate pairs on the two orthogonal polarizations.
At the receiver, the signals on the two polarizations are coherently
superimposed to cancel the first-order nonlinear distortions. 

Only a few studies consider the effects of the random polarization
impairment, such as polarization-dependent loss (PDL), on the performance
of the digital NLC techniques. In {[}7{]}, an investigation of the
impact of polarization effects on the performance of digital back
propagation and perturbation-based NLC is performed. However, to the
best of our knowledge, no investigation of the PDL impact on the DPC
technique is considered in the literature. 

In this letter, we investigate the impact of the PDL on the performance
of the recently proposed LPC-PCTS technique. The PDL-induced signal
power imbalance between the two polarizations disrupts the cross-correlation
property of the nonlinear impairments, which can affect the distortion
cancellation through the coherent superposition of the LPC-PCTS technique.
Therefore, we introduce a modified procedure, referred to as the linear
time-coded (LTC)-PCTS technique, where we transmit linearly coded
phase conjugate pairs on adjacent time slots of the same polarization,
and compare its PDL tolerance with that of the LPC-PCTS technique.
We carry out the investigation with both aligned- and statistical-PDL
models. We present results to demonstrate that the LTC-PCTS shows
a superior PDL tolerance when compared to the LPC-PCTS, regardless
of the PDL model. 

The remainder of this letter is organized as follows: Section II describes
the aligned- and statistical-PDL models, and Section III explains
the simulation setup. In Sections IV and V, the impact of PDL on the
two DPC approaches is compared, by considering the aligned- and statistical-PDL
models, respectively. Section VI draws the conclusions. \vspace{-0.05cm}

\begin{figure*}[t]
\begin{centering}
\includegraphics[width=0.9\textwidth,height=0.16\paperheight]{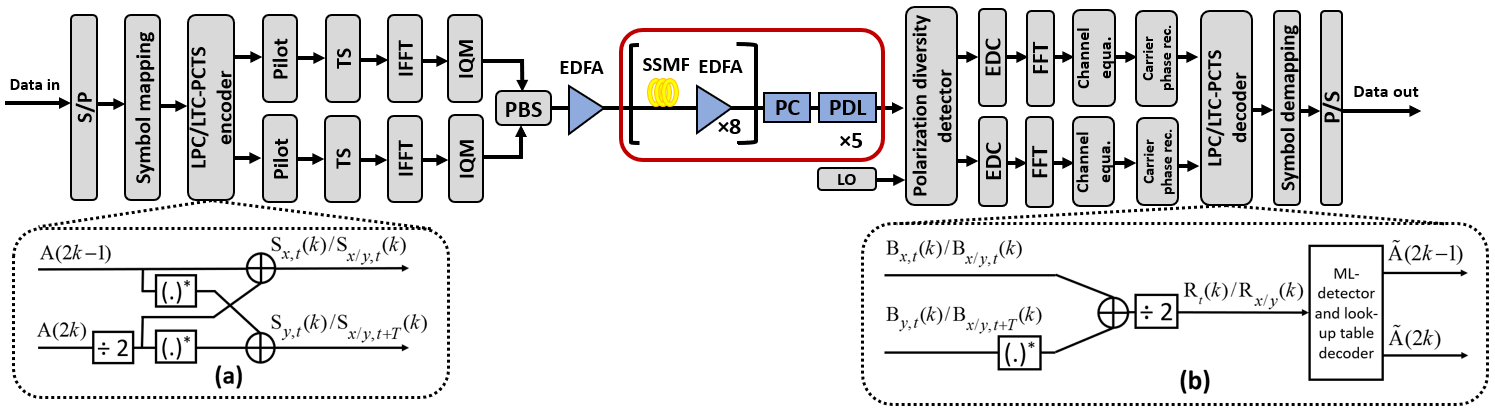}
\par\end{centering}
{\scriptsize{}Fig. 2: The simulation setup for one channel with a
5-section PDL emulator along the link. Insets (a) and (b) show the
encoder and decoder of the LPC/LTC-PCTS techniques, respectively.
$A$ and $S$ represent the OFDM symbols before and after the encoder;
$B$ and $R$ are the OFDM symbols before and after the coherent superposition;
and $\tilde{A}$ represents the recovered symbol after the decoder.
$x$/$y,$ $t$ and $T$ represent the horizontal/vertical polarization,
time and symbol duration, respectively. $k=1,2,...,K/2$, where $K$
is the subcarrier number and \textasteriskcentered{} stands for the
complex conjugation operation. S/P: serial-to-parallel, TS: training
symbol, (I)FFT: (inverse) fast Fourier transform,  IQM: inphase/quadrature
phase modulator, PBS: polarization beam splitter, EDFA: erbium doped
fiber amplifier, SSMF: standard single mode fiber, PC: polarization
controller, LO: local oscillator, EDC: electronic dispersion compensation,
ML: maximum-likelihood, P/S: parallel-to-serial.}{\scriptsize \par}
\end{figure*}
\vspace{-0.25cm}

\section{The Aligned- and Statistical-PDL Models}

PDL is caused by the polarization dependence on the transmission properties
of optical components, where one polarization component of the signal
suffers more loss than the other. The input/output field relation
of a PDL element, rotated with respect to the signal state-of-polarization
(SOP) by an angle $\theta,$ can be expressed using the Jones matrix
representation as {[}9{]}:

\begin{equation}
\left[\begin{array}{c}
u_{x}(t)\\
u_{y}(t)
\end{array}\right]=\left[\begin{array}{cc}
\cos\theta & -\sin\theta\\
\sin\theta & \cos\theta
\end{array}\right]\left[\begin{array}{cc}
1 & 0\\
0 & \alpha
\end{array}\right]\\
\left[\begin{array}{cc}
\cos\theta & \sin\theta\\
-\sin\theta & \cos\theta
\end{array}\right]\left[\begin{array}{c}
v_{x}(t)\\
v_{y}(t)
\end{array}\right],
\end{equation}
where $[v_{x}(t)\,\,v_{y}(t)]^{\dagger}$ and $[u_{x}(t)\,\,u_{y}(t)]^{\dagger}$
represent the input and output optical fields, respectively, with
the superscript $\dagger$ as the transpose. The parameter $0<\alpha<1$
is the PDL coefficient defined as the ratio between the minimum and
maximum transmission intensities; this is related to the PDL measured
in decibels (dB) as $\rho=-20\,\textrm{log}\,\alpha$ {[}10{]}. \vspace{0.25cm}

The PDL impact has been studied in coherent optical systems using
two different models: the aligned- and statistical-PDL models {[}7,11{]}.
In the aligned-PDL model, the signal SOP and the PDL axes of the optical
components are aligned with the same rotation angle $\theta.$ Fig.
1 (a)-(b) shows two cases for the aligned-PDL model with $\theta=0^{0}$
and $45^{0},$ respectively. At the rotation angle $\theta=0^{0},$
the optical signal-to-noise ratio (OSNR) of one of the polarizations
is degraded when compared to the other. On the other hand, for $\theta=45^{0},$
the PDL causes same OSNR degradation for both polarization components
along with the signal cross-talk due to the loss of orthogonality.
In {[}11{]}, it has been shown that pathological cases of the aligned-PDL
elements, such as $\theta=0^{0}$ and $45^{0},$ are the worst cases
of PDL in linear and nonlinear regimes, respectively.

\vspace{-0.25cm}

\begin{figure}[H]
\begin{centering}
\includegraphics[width=0.6\columnwidth,height=0.34\paperheight]{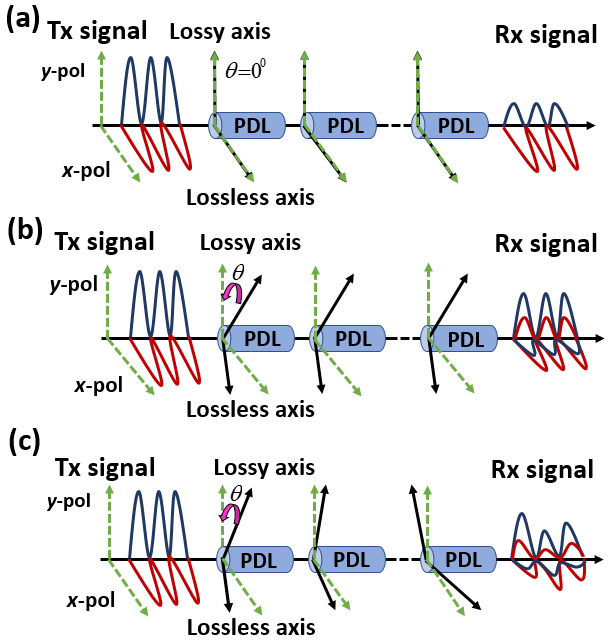}
\par\end{centering}
{\scriptsize{}Fig. 1: Schematic diagram of the impact of PDL on the
PDM signal. (a) aligned-PDL with $\theta=0^{0},$ (b) aligned-PDL
with $\theta=45^{0},$ and (c) statistical-PDL with random rotation
angle $\theta.$ }{\scriptsize \par}
\end{figure}

\vspace{-0.25cm}

In the statistical-PDL model, the rotation angle $\theta$ varies
uniformly within $[0,2\pi),$ as shown in Fig. 1 (c). This induces
the random signal power and OSNR fluctuations between the two polarizations.
In this case, the total cumulated PDL has a Maxwellian distribution
with the root mean square (rms) value $\rho_{rms}=\rho\sqrt{N},$
where $N$ is the number of spans {[}1{]}. \vspace{-0.1cm}

\section{Simulation Setup }

The simulation setup used to study the PDL impact on the DPC approaches
is shown in Fig. 2. We consider a $5$-section PDL emulator, which
closely approximates a real system {[}7{]}. In this setup, the signal
interacts with the PDL element after propagating through eight spans
of standard single mode fiber (SSMF). Five such loops realize a $5$-section
PDL emulator.\textcolor{black}{{} The PDL along the transmission link
mainly comes from the lumped optical elements. In a realistic transmission
link, such optical elements are placed after several fiber spans.
Therefore, placing a PDL element after eight spans of fiber is sufficient
enough to study its impact on the performances of the DPC approaches
{[}7{]}.} A polarization controller is placed before the PDL element
to control the signal SOP after each round trip. Insets (a) and (b)
show the encoder and decoder, respectively, for both LPC/LTC-PCTS
techniques. The encoder linearly combines the data symbols on the
adjacent subcarriers of the OFDM symbol, one at full amplitude and
the other at half amplitude. \textcolor{black}{This linear coding
provides an equal distance between the constellation points and reduces
the average probability of symbol error {[}6{]}. The linearly combined
symbols are then transmitted as phase conjugate pairs on the orthogonal
polarizations/time slots. At the receiver, after coherent superposition,
maximum-likelihood detection of the recovered symbol is carried out
on a symbol-by-symbol basis and a look-up table decoder is applied
to obtain the transmitted data symbols {[}6{]}.}\vspace{0.1cm}

The transmission system consists of a wavelength division multiplexed
(WDM) CO-OFDM superchannel employing the LPC/LTC-PCTS techniques.
The superchannel comprises four OFDM sub-bands with a frequency spacing
of $37.5$ GHz. The baud rate is $32$ Gbaud. The OFDM symbol consists
of $3300$ data subcarriers, and the inverse fast Fourier transform
size is $4096$ {[}6{]}. In each OFDM symbol, four pilot subcarriers
are inserted for the common phase error compensation and a cyclic
prefix of $3$\% is added. \textcolor{black}{The modulation format
is quadrature phase-shift-keying (QPSK). }Therefore, the net data
rate is $401.33$ Gb/s. The long-haul fiber link consists of $40$
spans of SSMF, each having a length of $80$ km, an attenuation coefficient
of $0.2$ dB/km, a nonlinear parameter of $1.22$/(W.km), and a dispersion
parameter of $16$ ps/nm/km. The optical power loss for each span
is compensated by an erbium doped fiber amplifier with $16$ dB gain
and $5.5$ dB noise figure. The transmitter and receiver lasers have
the same linewidth of $100$ kHz. At the receiver, after the polarization
diversity detector, the dispersion compensation is carried out using
the overlapped frequency domain equalizer with the overlap-and-save
algorithm {[}12{]}. The channel equalization and carrier phase recovery
are carried out as in {[}13{]}. \vspace{-0.1cm}

\section{Performance Evaluation with Aligned-PDL }

The DPC is a generalized technique in which one can use orthogonal
polarization states or time slots to transmit the phase conjugate
pairs {[}5{]}.The motivation behind the recently proposed LPC-PCTS
technique is to solve the issue of halving the spectral efficiency
associated with the PCTW technique {[}6{]}. Its effectiveness strongly
depends on the cross-correlation between the nonlinear distortions
added onto the transmitted phase conjugate pairs on the two polarizations.
However, the polarization cross-talk induced signal power imbalance
between the two polarizations may disrupt this cross-correlation property
and significantly degrade the performance of the LPC-PCTS technique.
For this reason, we consider the LTC-PCTS approach, in which we transmit
the linearly coded phase conjugate pairs on adjacent time slots of
the same polarization.

\textcolor{black}{In Fig. 3, we consider two simulation scenarios
to investigate the performances of the DPC approaches: one is with
PDL alone and the other is with the polarization mode dispersion (PMD)
and PDL. In both cases, the performances are showed for the two worst
case aligned-PDL scenarios with the rotation angles} $\theta=0^{0}$
and $45^{0},$ respectively. The optical launch power is set at the
optimum value of $-3$ dBm per channel. When considering only the
PDL, the performance of the LPC-PCTS technique with $\theta=45^{0}$
monotonically decreases as the PDL value increases. The LTC-PCTS technique
shows an improved performance, above the soft-decision forward error
correction (SD-FEC) limit for the two considered worst case scenarios
of the aligned-PDL. The PMD effect is included in the transmission
fiber by choosing a typical mean differential group delay of $20$
ps, as in {[}14{]}. It is observed that the performances of the DPC
approaches are significantly affected for the case considering both
PMD and PDL. The interplay between PMD and PDL distorts a communication
system more than either effects alone. A separate study is required
to understand the interaction between PMD and PDL in the optical transmission
link. Henceforth, we focus only on the impact of PDL on the performances
of the DPC approaches. \vspace{-0.25cm}

\begin{figure}[H]
\begin{centering}
\includegraphics[width=0.75\columnwidth,height=0.32\paperheight]{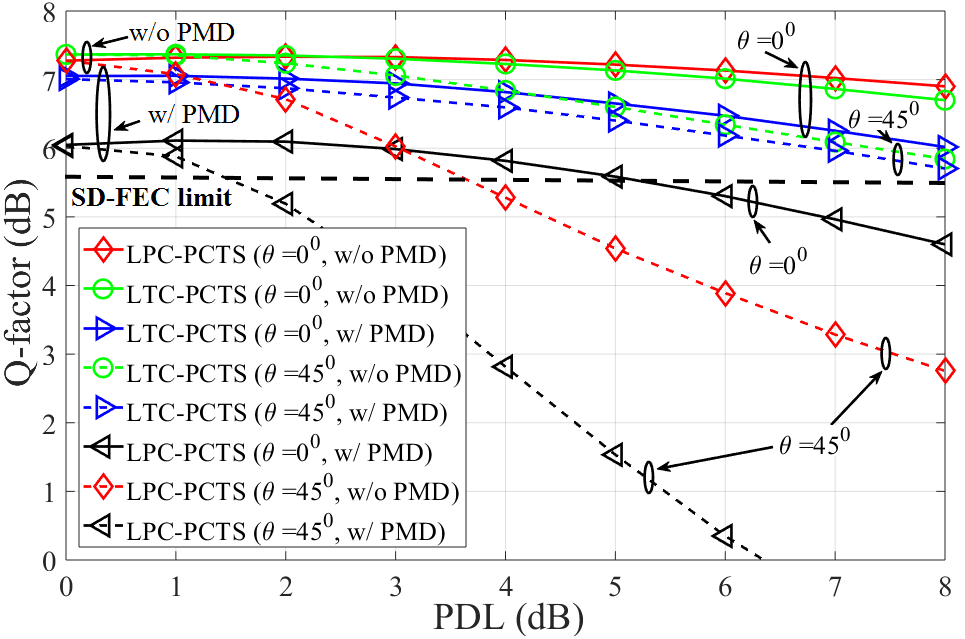}
\par\end{centering}
{\scriptsize{}Fig. 3: $Q$-factor as a function of PMD and PDL with
rotation angles $\theta=0^{0}$ and $45^{0}$ at the optimum launch
power of $-3$ dBm. SD-FEC: soft-decision forward error correction.}{\scriptsize \par}
\end{figure}

\vspace{-0.25cm}
In Fig. 4, the $Q$-factor performance of the DPC approaches is shown
for different values of the fiber launch power at a fixed aligned-PDL.
In this case, we select a PDL value of $3.6$ dB. This corresponds
to the rms value of the cumulated PDL in the evaluation with statistical-PDL
model, as it will be discussed in Section V. We start analyzing the
linear regime, i.e, the initial increasing part of the $Q$-factor
curves. For both DPC approaches, the performance in the presence of
PDL when $\theta=0^{0}$ is lower when compared to the absence of
PDL. This degradation in performance can be explained by the impact
of the PDL-induced OSNR imbalance between the two polarizations. For
$\theta=45^{0},$ the performance of the LPC-PCTS is significantly
reduced when compared to the LTC-PCTS. This is due to the signal cross-talk
induced power fluctuations on the two polarizations along with the
OSNR degradation. At high input powers, where the performance is limited
by the nonlinear distortions, the DPC approaches with $\theta=0^{0}$
perform slightly better than the case without PDL. This can be explained
by the decrease of the higher-order nonlinear distortions. In fact,
in the presence of PDL, the signal in one polarization is attenuated
more than in the other. This leads to the reduction of higher-order
nonlinear distortions after the coherent superposition. Note that
the higher-order nonlinear distortions are not canceled by the DPC
approaches {[}5{]}. It is observed that the performance of the LPC-PCTS
technique with $\theta=45^{0}$ degrades when compared to the LTC-PCTS
technique in both linear and nonlinear regimes. This is because the
polarization cross-talk, due to the loss of orthogonality, causes
signal power fluctuations on the two polarizations {[}8{]}. \vspace{-0.25cm}

\begin{figure}[H]
\begin{centering}
\includegraphics[width=0.7\columnwidth,height=0.34\paperheight]{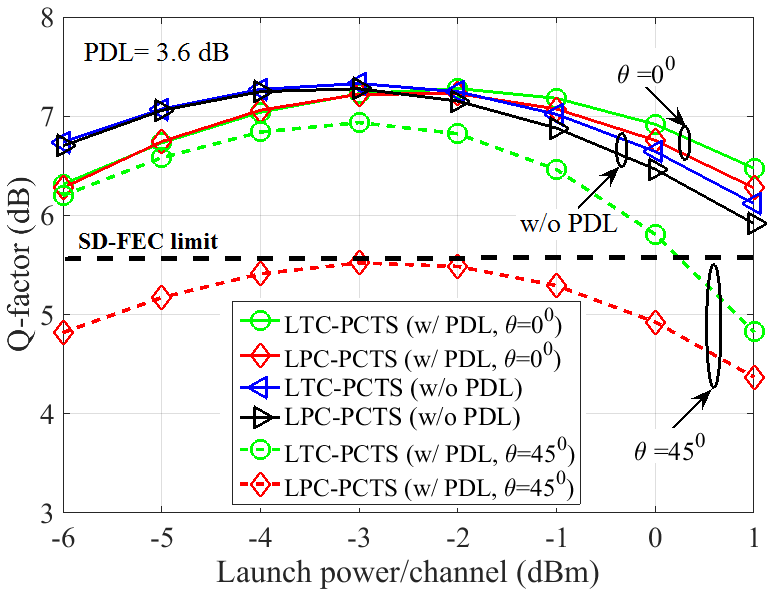}
\par\end{centering}
{\scriptsize{}Fig. 4: $Q$-factor as a function of launch power in
the presence and absence of the PDL.}{\scriptsize \par}
\end{figure}
\vspace{-0.25cm}

\begin{figure}[H]
\begin{centering}
\includegraphics[width=0.7\columnwidth,height=0.34\paperheight]{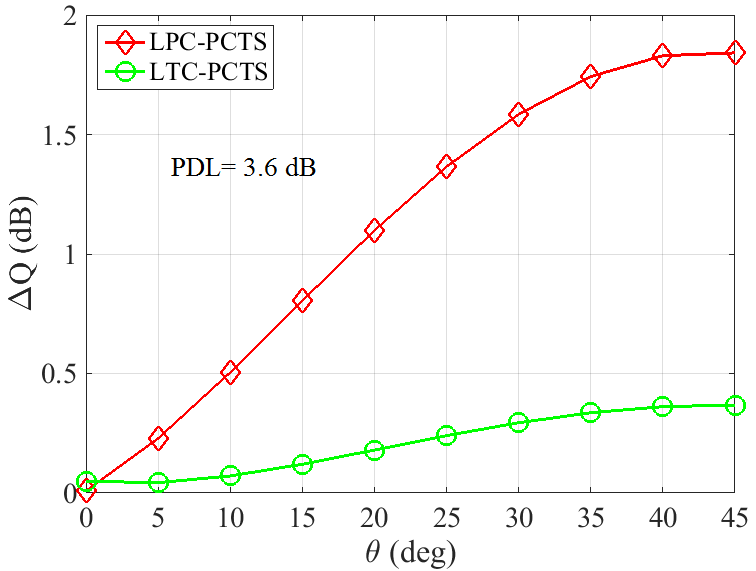}
\par\end{centering}
{\scriptsize{}Fig. 5: $Q$-factor penalty $(\Delta Q)$ for different
rotation angles, $\theta.$ The optical launch power is fixed at the
optimum value of $-3$ dBm.}{\scriptsize \par}
\end{figure}
\vspace{-0.25cm}

In Fig. 5, the performance of both DPC approaches is presented as
a function of the rotation angle $\theta$ in terms of the $Q$-factor
penalty. The $Q$-factor penalty is defined as: $\Delta Q=Q_{opt}-Q,$
where $Q_{opt}$ is the $Q$-factor at the optimum launch power when
PDL is not considered. It is seen that the $Q$-factor penalty is
maximum at the rotation angle $\theta=45^{0}$ and minimum at $\theta=0^{0}.$
We also observe that the cross-talk induced $Q$-factor penalty at
$\theta=45^{0}$ for the LPC-PCTS technique is $1.84$ dB, while it
is only about $0.35$ dB for the LTC-PCTS technique. It should be
noted that further increasing the angle from $\theta=45^{0}$ to $90^{0}$
would result in the mirror image of the plot with a minimum $Q$-factor
penalty at $\theta=90{}^{0}$ because of the lower cross-talk induced
power fluctuations. Therefore, we provide results for angles ranging
from $\theta=0^{0}$ to $45^{0}$ only. \vspace{-0.25cm}

\section{Performance Evaluation with Statistical-PDL }

The $Q$-factor distribution presents a more realistic impact of the
PDL on the performances of DPC approaches. Fig. 6 shows the estimated
\textcolor{black}{probability density function (PDF)} of the $Q$-factor
in the presence and absence of PDL for both LPC/LTC-PCTS techniques.
We carried out Monte Carlo simulation to estimate the $Q$-factor
PDF by using $500$ random seeds of the signal SOP and the PDL orientation
angle $\theta$ in the limit $[0,2\pi).$ We select a typical PDL
value of $\rho=1.6$ dB {[}7{]}. This gives an rms cumulated PDL value
for a $5$-section PDL emulator of $3.6$ dB (i.e., $1.6*\sqrt{5}$).
The fiber launch power is fixed at the optimum value of $-3$ dBm.\vspace{-0.25cm}

\begin{figure}[H]
\begin{centering}
\includegraphics[width=0.7\columnwidth,height=0.34\paperheight]{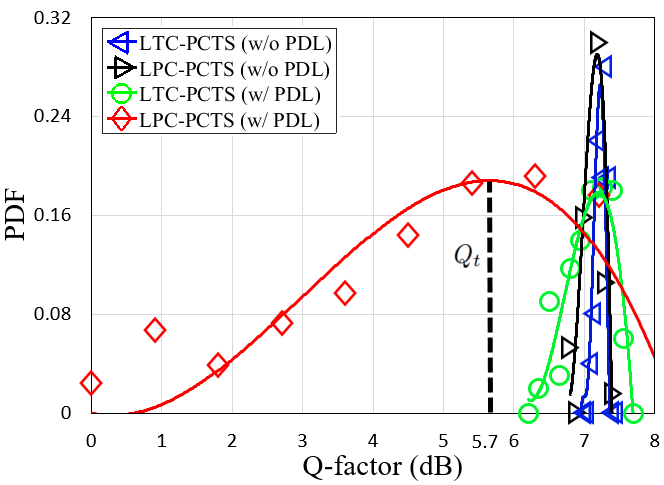}
\par\end{centering}
{\scriptsize{}Fig. 6: $Q$-factor PDF for LPC-PCTS and LTC-PCTS at
$\rho_{rms}=3.6$ dB and optical launch power = $-3$ dBm. }{\scriptsize \par}
\end{figure}

\vspace{-0.25cm}
The results indicate that without PDL, the $Q$-factor distributions
for both DPC approaches are very narrow. On the other hand, in the
presence of PDL, the $Q$-factor distribution of LPC-PCTS significantly
enlarges, which leads to an increased outage probability. We define
the outage probability as the probability that the random $Q$-factor
is less than a certain threshold value, i.e., $\mathfrak{\mathscr{\mathbb{P}}}\textrm{r}[Q<Q_{t}],$
where $Q_{t}$ is the threshold. For instance, assuming a $Q_{t}$
value of $5.7$ dB corresponding to the SD-FEC limit {[}15{]}, the
outage probability for LPC-PCTS in the presence of PDL is $0.63.$
However, it is observed that the outage probability for LTC-PCTS approaches
zero in the presence of PDL. This indicates that the approach which
uses the orthogonal time slots of the same polarization is only slightly
affected by the PDL-induced distortions. 

\section{Conclusion}

We investigated the performance penalties induced by the PDL on the
LPC/LTC-PCTS techniques, in a CO-OFDM superchannel system. We carried
out a comprehensive numerical study using aligned- and statistical-PDL
models. In the investigation with the aligned-PDL, LTC-PCTS shows
a superior PDL tolerance when compared to LPC-PCTS. The $Q$-factor
performance of the former is above the SD-FEC limit for the pathological
cases of all aligned-PDL with $\theta=0^{0}$ and $45^{0}.$ The investigation
with the statistical-PDL model also indicates that LTC-PCTS outperforms
the LPC-PCTS technique, with the former providing an outage probability
approaching zero for an rms PDL value of $3.6$ dB. We concluded that
while LPC-PCTS is severely affected by the PDL-induced distortions,
LTC-PCTS still provides a good performance under such conditions.
In future, this work can be extended to consider the impact of the
interplay between PMD and PDL on the performance of the LPC/LTC-PCTS
techniques. \vspace{-0.1cm}

\section*{Acknowledgement} 

This work was supported in part by Atlantic Canada Opportunities Agency
(ACOA) and Research Development Corporation (RDC), NL, Canada.

\end{document}